# GSM-GPRS Based Smart Street Light

Imran Kabir[1], Shihab Uddin Ahamad[2], Mohammad Naim Uddin[3], Shah Mohazzem Hossain[4], Faija Farjana[5], Partha Protim Datta[6], Md. Raduanul Alam Riad[7], Mohammed Hossam-E-Haider[8]
Military Institute of Science and Technology (MIST), Dhaka-1216, Bangladesh
imran1996@gmail.com; sabbir.shihab@gmail.com; naim_shekh@hotmail.com; mohazzem_hossain@eece.mist.ac.bd; faija@eece.mist.ac.bd;
partha.datta@ieee.org; raduanulalamriad130749@gmail.com; haider8400@yahoo.com

***Abstract*-Street lighting system has always been the traditional manual system of illuminating the streets in Bangladesh, where a dedicated person is posted only to control the street lights of a zone, who roams around the zonal area to switch on and switch off the lights two times a day, which brings about the exhibition of bright lights in street even after sunrise and in some cases maybe the whole day. This results in insertion to the budget. In addition to this, faulty lights may not come to the heed of the concerned authority for a long time which leads to the technical downside. This paper demonstrates a process of controlling the street lights in country like Bangladesh employing SIM900 GSM-GPRS Shield which comes up with the provision of manual control, semi-automated control as well as full-automated control.**

## I. Introduction

Around 5% of total energy is consumed by the street lights, lighting of parking lots, pedestrian area along with park lighting [1] which contains both the working part and the wasted part. The wastage of the power occurs mainly due to the timing of manual operation of lights. This paper is a projection of a process which demolishes the wastage in case of Bangladesh. A practical implementation of this model will also be presented which took place in Mirpur Cantonment, Dhaka, Bangladesh.

This model is structured using the GSM-GPRS shield. The whole model has the superiority to be controlled in full-automated, semi-automated and manual method. The GPRS part has the access to internet which can use the sunset and sunrise timing to allow the system to operate in full-automated method. On the other hand, utilizing the GSM part and RTC module, semi-automated control can be administrated. Here, either a particular set of time can be placed to switch on-off the system or just sending a mobile message can execute the same. Even if there is issue with these automated systems, there is also a feature of manual control method. Moreover, the proposed system has the option of detecting the fault associated with potential transformer and current transformer. The entire process is controlled by ATmega-328p microcontroller.

## II. Literature Review

Suseendran et al. [2] represented the brightness controlling of the street light using sensors, based on IoT (Internet of Things), video vehicle detection and LDR (Light Decreasing Resistance) sensor. Each lamp unit encompassed two sensors- video vehicle detection sensor and LDR sensor. All the data were collected and processed on a regular basis which had required a huge data processing system.

Lavric et al. [3] revealed a practical implementation of street light monitoring and controlling system employing WSN (Wireless Sensor Network System). The authors had focused on the software-based method and then performed a real-time implementation using limited number of lights. This archetype included Doppler sensor to allow vehicle detection. According to the appearance of vehicle, lights for that particular region increased intensity.

Archibong et al. [4] worked with IoT based PV solar self-powered lighting system for street with anti-vandalism monitoring and tracking competency. LDR sensor switched the lights on-off along with an IR (InfraRed) sensor having the ability to save power. The anti-vandalism system was installed with a User Interface (UI), where the street lights communicated with the people and devices through wi-fi module at the control station.

Prasad et al. [5] projected a case-study of Nagpur street lighting system, where LED lights were utilized along with motion detection system. The arbitration showed that energy consumption was lessen by 55% per month.

Abdullah et al. [6] worked to provide a suitable system to minimize the power dissipation. The prototype comprised LDR, IR, battery and LED along with Arduino uno, which was skilled in increasing the intensity of light based on the speed of the vehicle. This project concluded that, this system will retain about 40% energy per month.

Smart street lighting system also practiced in another way by Sukhathai et al. [7] using LoRaWAN in accompany with motion detector and illuminance sensor to mitigate the power consumption. In this model, every street light had been furnished with LoRaWAN communication module, which controlled the LED lights by exchanging data with main server.

Mary et al. [8] investigated on the control of street lights using lamp unit, sensor unit and access unit. The lamp and the sensor units accommodated sensors for identifying motion and brightness of controller, LED device and communication device. For the access unit, Zigbee was used. Whenever there was any motion, lights' brightness was proliferated otherwise it was dimmed or low.

All the above divulged papers are prototype based. Only the work with WSN network system represented a field work with implementing the concept only for controlling four lighting units. This paper portrays a practically implemented method with control of 300 lights of 2.5km distance.

## III. SYSTEM DESIGN

Fig. 1 incarnates the system control flow of the system. System design embraces two concerns- components used for the system device and simulation of the system.

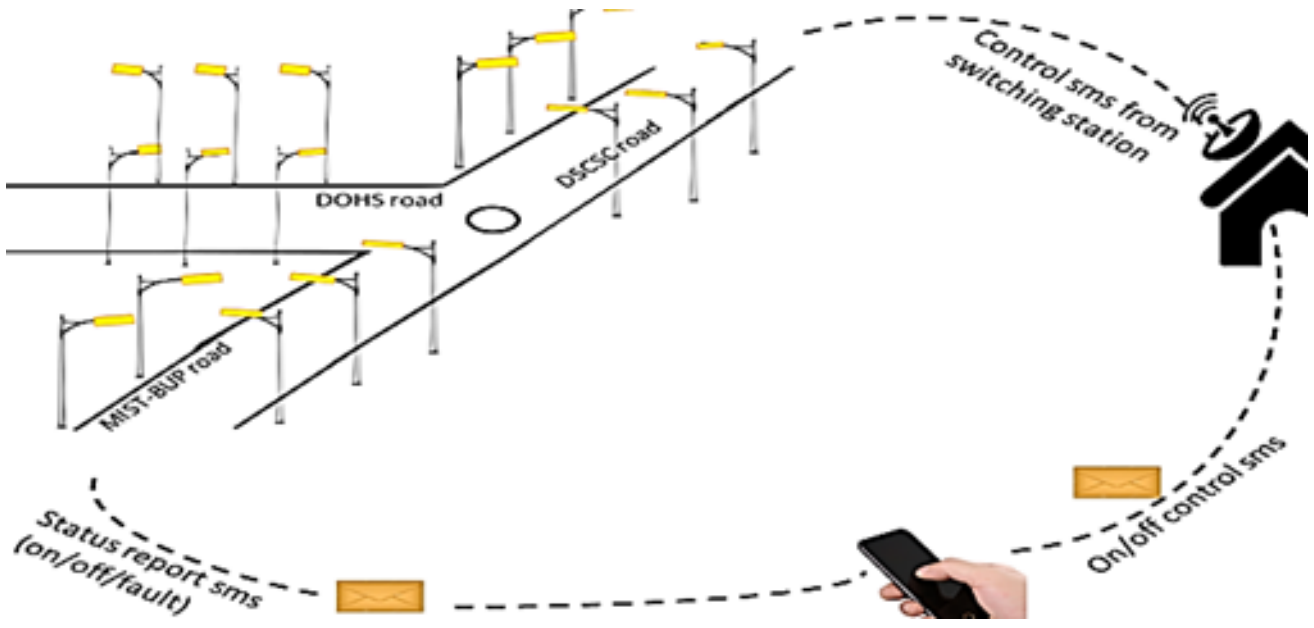

Fig.1. System control flow

### A. Components

The system owns four leading in addition with some other ancillary constituents, authorizing the real time control.

- ***ATmega-328p:*** ATmega-328p possesses 8-bit AVR RISC processor core [9]. This is the heart component of the proposed system with the sanction of controlling the messaging process, RTC module, relay, real-time data acquirement of sunrise-sunset along with taking input from keypad, power flow data processing and decision making.
- ***GSM-GPRS Shield:*** GSM-GPRS shield comprises both GSM (Global System for Mobile communication) and GPRS (General Packet Radio Service). The GSM module is used for mobile messaging method. On the other hand, the GPRS module is used for connecting the internet service.
- ***Clock Module****:* DS1302 RTC (Real Time Clock) module is accumulated with the system device to set the desired time for switching the lights on and off.
- ***PCR8574 IC:*** This I2C converter is deployed to act as an adapter to ameliorate the number of ports of the microcontroller while connecting to the display.
- ***Current and potential transformer:*** To concede the fault detection characteristics of the system, SCT013 clamp current transformer and 230:12 iron core potential transformer is adhered. They measure the power flowing through the structure and notify the device if there is any depletion from the threshold level. The dilution of power indicates the fault and then a message will be sent to the authority manifesting the issue.
- ***GSM antenna:*** Even though there is already an antenna inside the GSM-GPRS shield, additional antennas are accompanying the system to boost the signals.
- ***High Power MOSFET:*** IRF540N MOSFET, assigned in the circuit to work as a switching device for passing the signal from the micro-controller to the relay.
- ***Relay:*** 5V and 12V relays are affixed to the system to act as switch as well as an isolator in between main power line and microcontroller.
- ***SIM:*** A SIM is implemented in inner architecture with the GSM-GPRS shield for mobile communication.

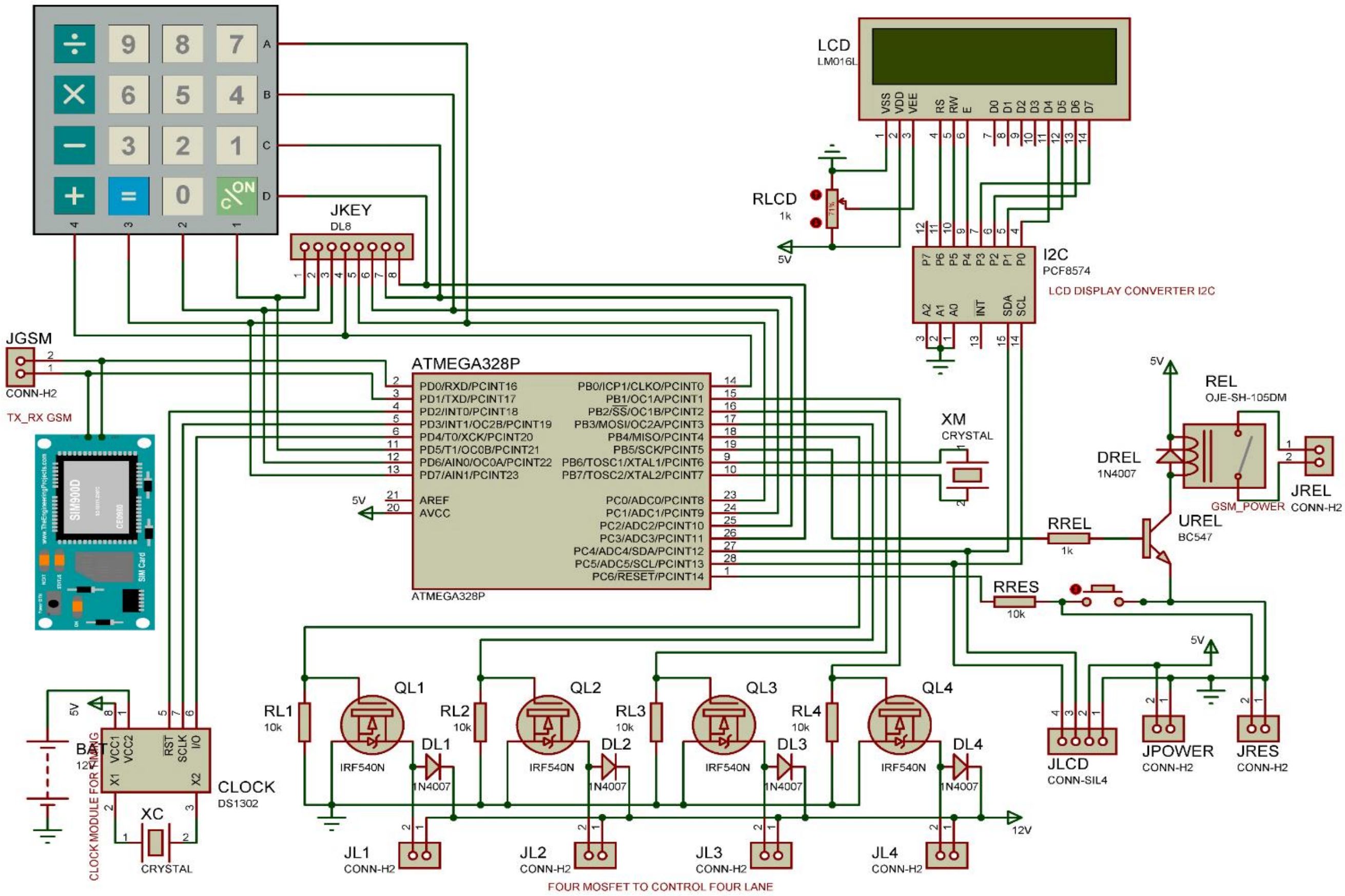

Fig. 2. Schematic diagram of prototype

- ***Keyboard:*** Matrix-keypad is anchored to the system device in order to regulate through cell phone network.
- ***LCD screen:*** 16:2 display is equated to the system device to evince the output of the system.

## B. Simulation

Fig. 2 symbolizes the schematic diagram of the prototype of system device. In the internal section of the system device, GSM-GPRS shield and keypad is used as input medium. 8 pins of the keypad indicate the row and column matrix. From the figure, A, B, C and D ports of keypad epitomize the row matrix where 1,2,3 and 4 ports indicate the column matrix. These 8 ports are coupled to JKEY which is internally linked with the microcontroller through PD5, PD6, PD7, PB0, PC0, PC1, PC2 and PC3 pins. PD5, PD6, PD7 and PB0 pins specify the columns 1, 2, 3 and 4 respectively. On the other hand, PC0, PC1, PC2 and PC3 pins stipulate the rows A, B, C and D separately.

For GSM-GPRS semi-automatic and full-automatic control, JGSM functions as a contactor to connect receiving and transmitting ports of the microcontroller for signaling and communicating data reciprocity which is connected to RXD and TXD pins of the microcontroller. JREL ports are fettered internally to the PB5 pin of the microcontroller for switching the GSM module on and off. GSM module confiscates a huge energy compared to the other units of the contrivance. To truncate this energy, the module can be turned off for a particular time by sending a message. The message contains a time duration for which the module will be turned off and the system will be in hibernation mode with extracting stunted energy.

The PD2, PD3 and PD4 pins of the microcontroller is hitched with the DS1302 clock module which maintains the pre-set timing of controlling the switching process. This module comes with a 32.768kHz crystal oscillator and on-board battery back-up [10]. The time for energizing and de-energizing the street lights is collimated by the keypad.

The commands of switching the lights are transmitted from PB1, PB2, PB3 and PB4 pins of the microcontroller. These pins are battened to four MOSFETs which are assigned as switches for the lane lights. Whenever any lane entails to be lighted up, the pin dedicated for that lane will acquire a high signal. In the MOSFET section, 12V is endowed for the prototype lights along with diodes to abrogate the reverse voltage.

16MHz extraneous oscillator is bridged to the XTAL1 and XTAL2 pins of the microcontroller to boost up the processing speed [11]. The 16:2 LCD display is attached to the SDA and SCL pins of microcontroller through PCR8574 I2C. This display obligates minimum 8 pins except the power pins to operate. Due to the shortage of pins in microcontroller, I2C is connected to it to intensify the number of ports. Later in the practical implementation, the outputs of the 4 MOSFETs are used to console the main street lights through magnetic contactor.

Fig. 3 incarnates the energy diagnosing process. Here SCT013 clamp 2000:1 current transformer estimates the current flow value which later converts into voltage and passes toward pin 19 of microcontroller named as ADC6 pin. Besides the 230:12 potential transformer measures the voltage quantity and then it passes the value to pin 22 of microcontroller termed as the ADC7 pin. The real time measurements of voltage and current are documented in the system. Whenever depleted value is ascertained, authority will be apprised about the reprehensible situation.

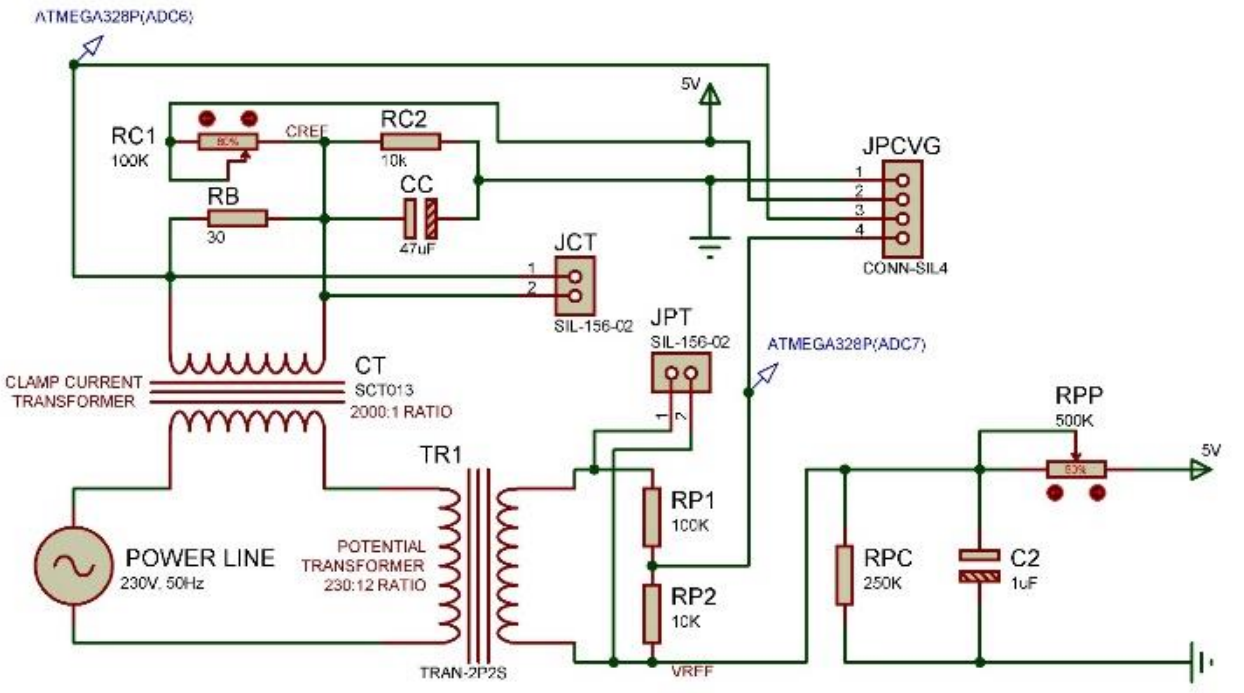

Fig. 3. Schematic diagram of energy meter

Fig. 4 demonstrates the practical implementation of the proposed system with four lane construction. The control process is done using the system device.

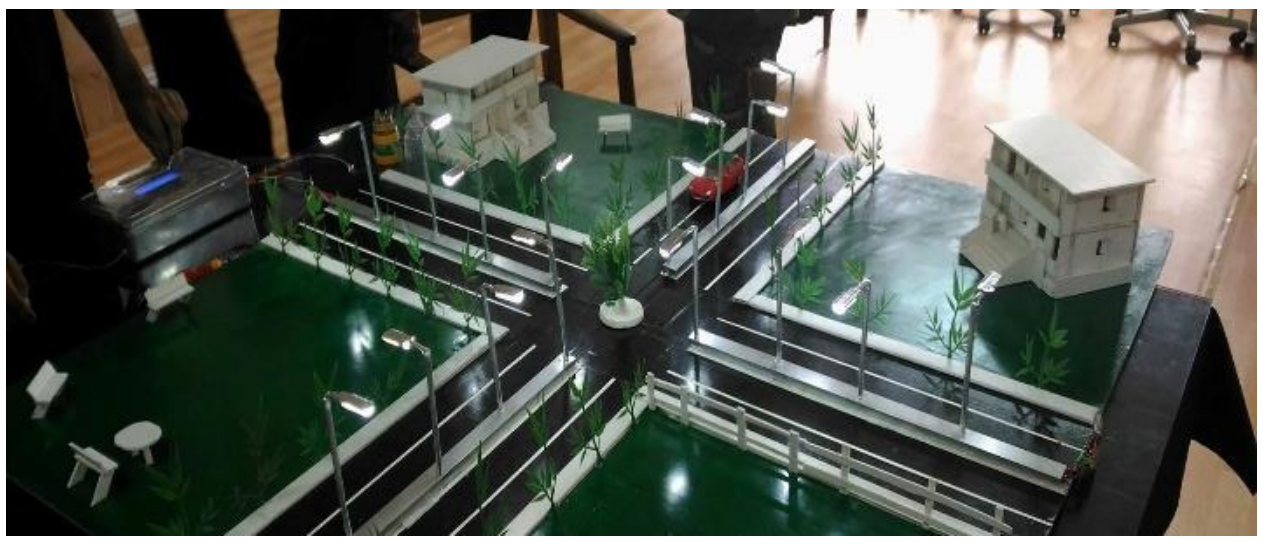

Fig. 4. Designed prototype

## IV. Implementation

After inheriting a successful prototype implementation, this model was installed in Mirpur Cantonment, Dhaka, Bangladesh for switching the street lights of Mirpur Cantonment area. This process results in substantiating a continuous service for 81 days without any technical barrier.

## A. PCB Design

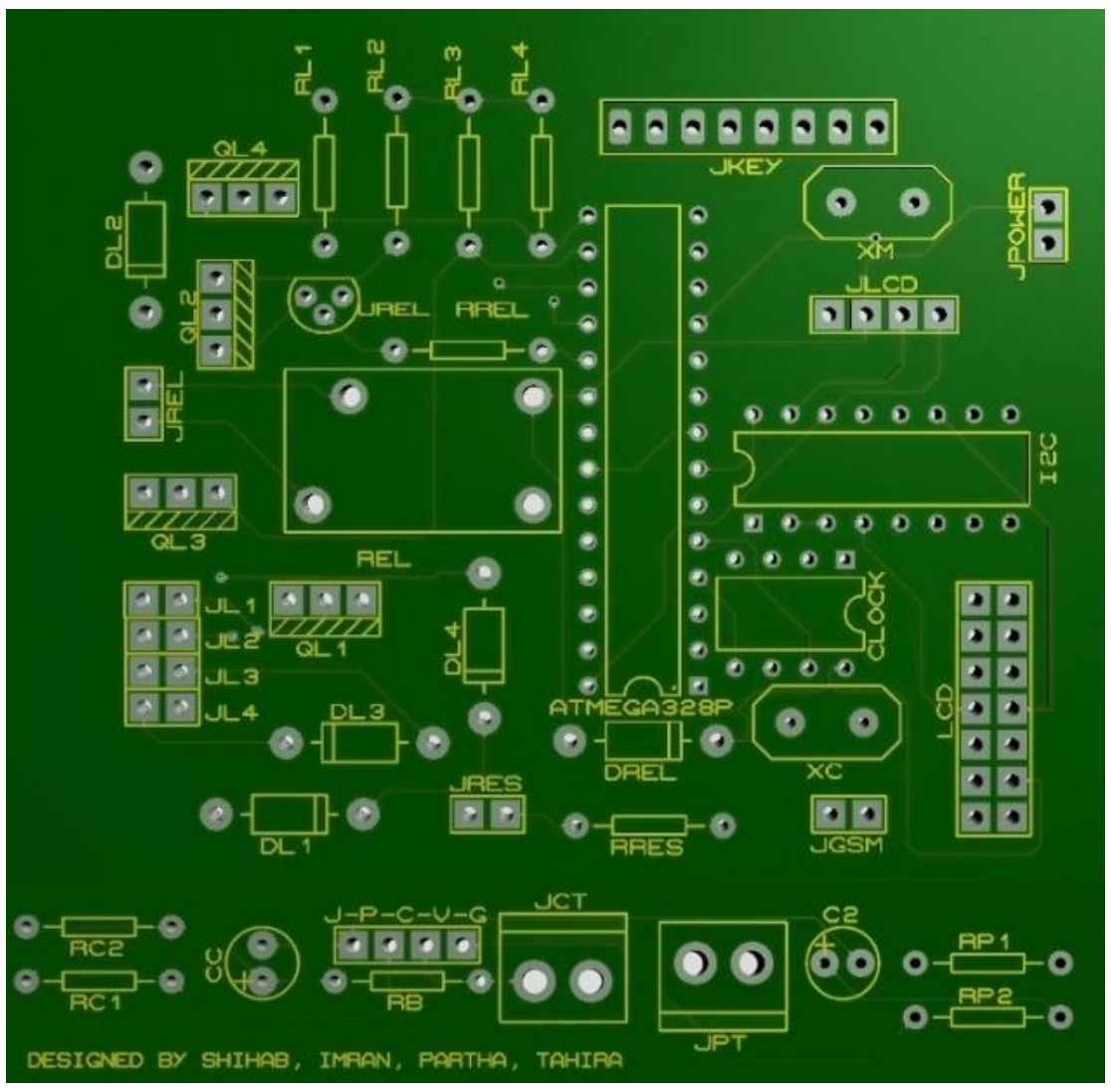

Fig. 5. PCB design of practical execution

PCB design, calls for the system device was executed using Proteus. Fig. 5 delineates the PCB design of the empirical application.

### B. Control Circuit

Fig. 6 represents the corporeal device of the control system including the LCD display, keypad and the power source.

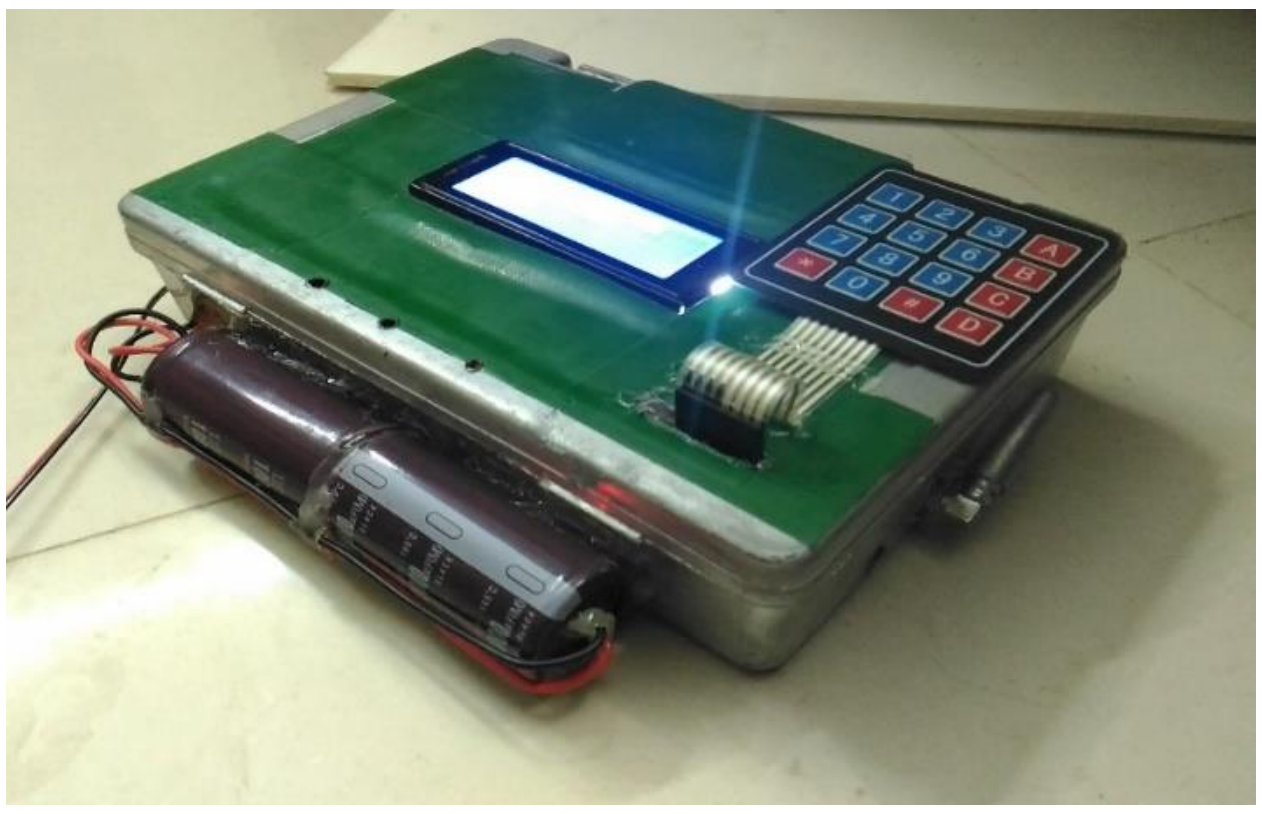

Fig. 6. Control device

Fig. 7 depicts the installed device in Mirpur Cantonment area switching station. This system works as a controller for 300 LED street lights covering 2.5km distance. Each light is of 25W.

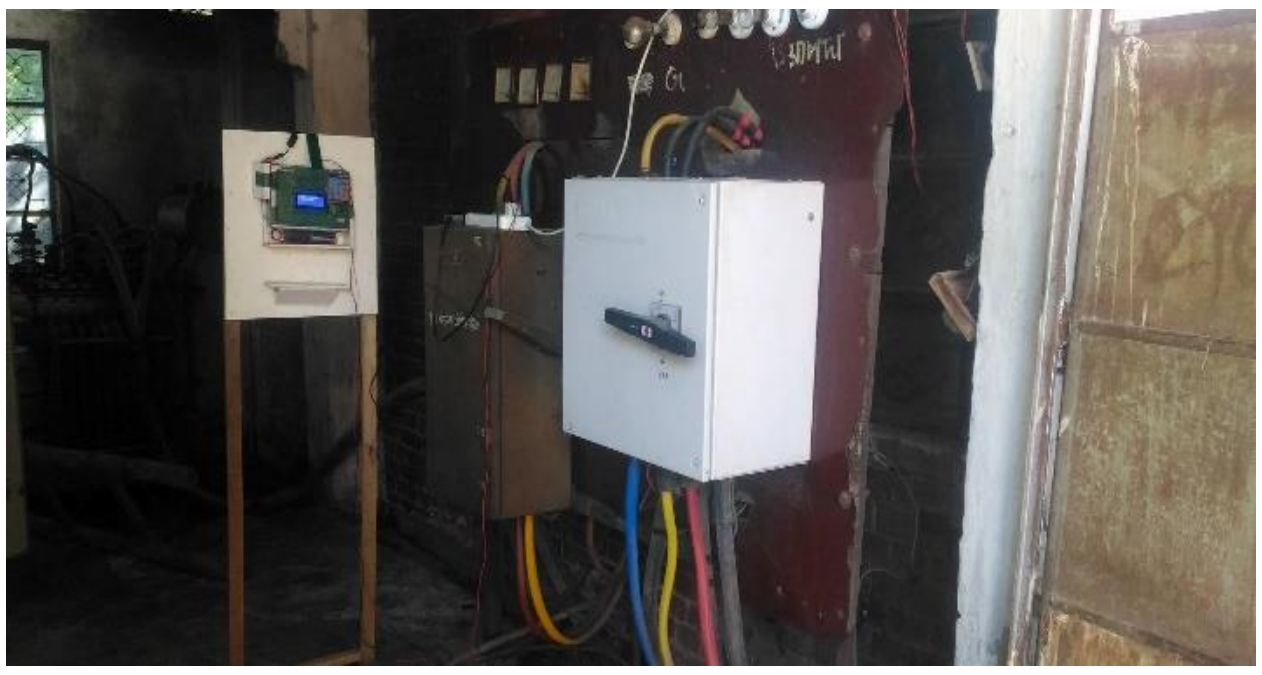

Fig. 7. Implementation in Mirpur Cantonment area switching station

Fig. 8 manifests the comparison in between the conventional and the proposed system where the red line shows the energy consumption with the conventional system on 22 July, 2019 on the other hand, the green line depicts the energy consumption with the proposed prototype on 24 July, 2019.

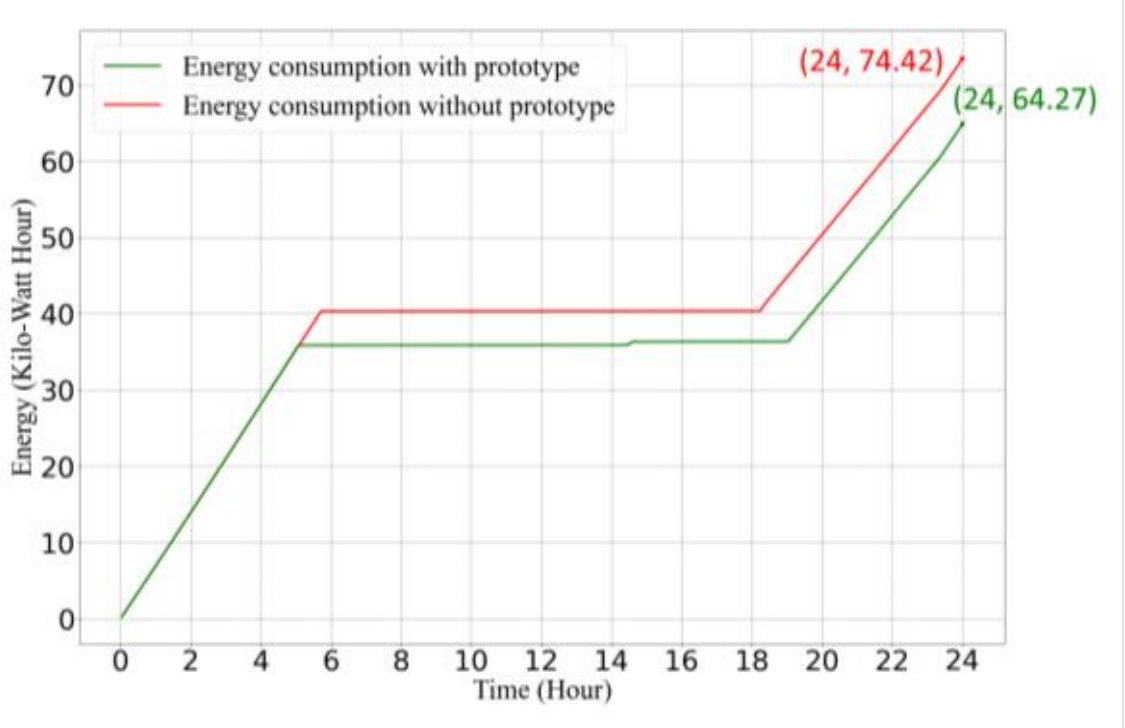

Fig. 8. Comparison between the conventional and the proposed system in case of energy

The power consumption for the two systems is shown in Fig. 9 where the green and the red lines are assumed to be the same as before. Sometimes after 14th hour, the proposed system is turned on for a very short moment to show if the lights were on, a large amount of energy can be consumed which will not happen in case of the proposed system.

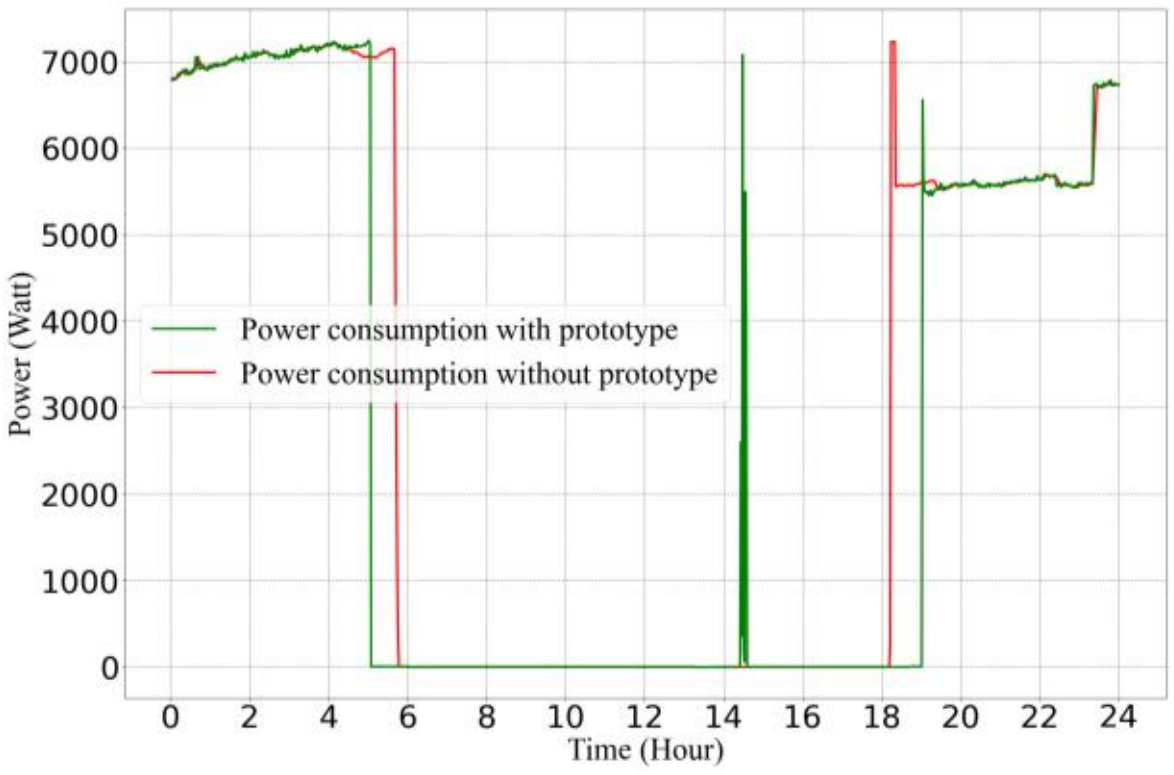

Fig. 9. Comparison between the conventional and the proposed system in terms of power consumption

## V. Developed Algorithm

"ARDUINO" is adopted to decipher the software part. For the GSM module, GSM library of Arduino is used. Input signal from the GSM enters to the microcontroller through RXD pin.

The device has the command to download the sunset time of a particular day and the sunrise time for the next day at 10:00AM of that day if the system is in full-automated mode. Otherwise, the normal mode, that is the semi-automated mode follows the DS1302 clock module to keep a track of previously placed timing.

Fig. 10 shows the flowchart of the system program algorithm. During the starting of the device, the display will show an interface to set the time. If the answer is positive, then times for date and time or sunset-sunrise or GSM sleep time is to be placed. On the other hand, for a negative command, the system will go to read and display the pre-set times. The microcontroller then read the GSM sunset-sunrise timing or the hibernation time, then the microcontroller will update its timing according to the command. If the message contains the on-off command for the lanes, microcontroller will send a high or low command to the pin dedicated for that lane and so control the switching system of the lanes.

Afterward, the microcontroller will go to read the current flow and voltage data from the fault detection part and calculate the value of power flow. If the value is subordinate to the threshold value, then a message will be sent to the concerned authority describing the issue otherwise the microcontroller will read the clock timing and compare it with sunset, sunrise and sleep time. According to the command, lights will be on or off.

After sending that command, system will check the device condition, either it is in on mode or in off mode. If the device is on, the whole process will continue otherwise the program will stop working.

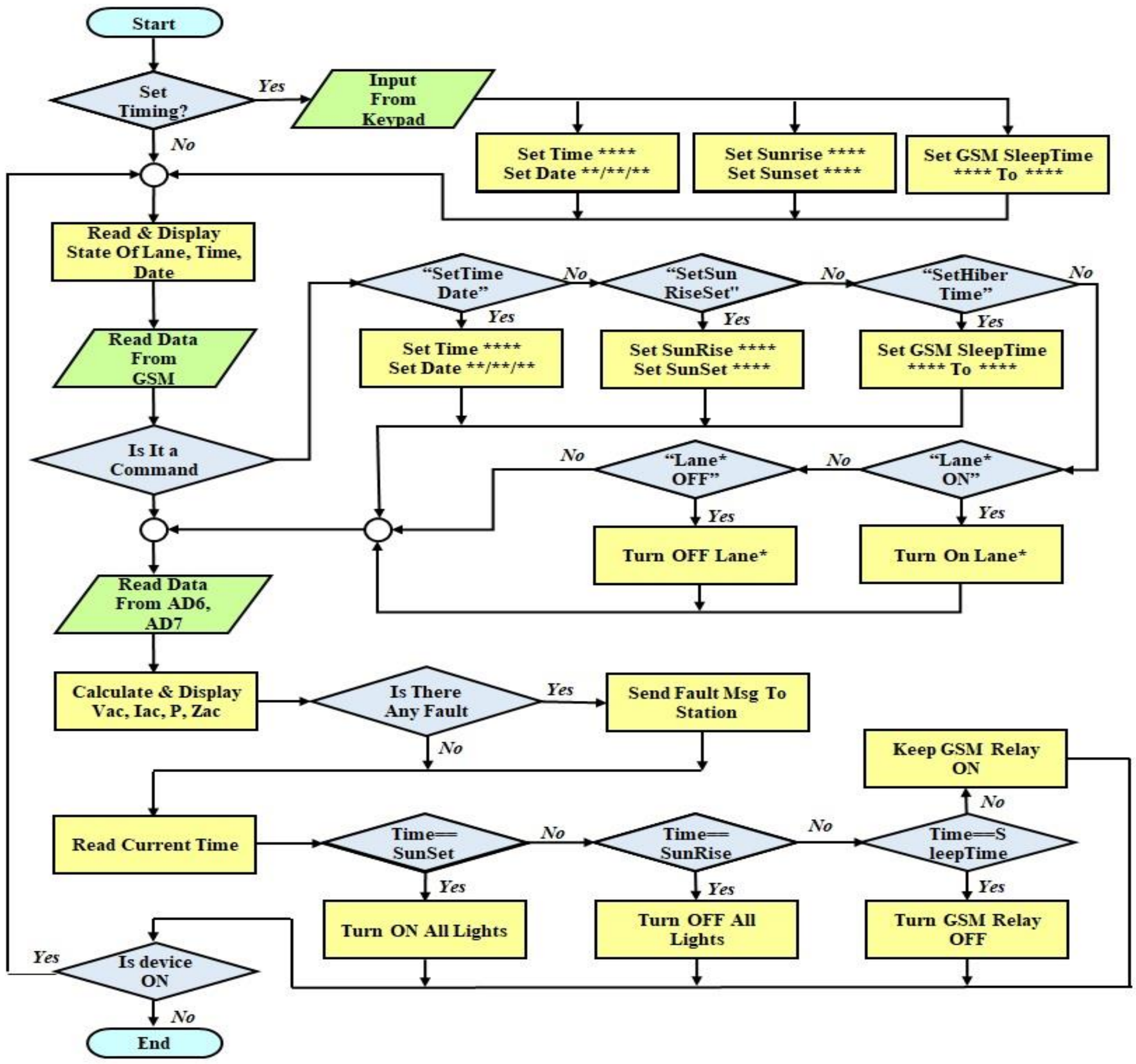

Fig. 10. Developed program algorithm

## VI. Conclution

The proposed system, based on GSM-GPRS shield is able to switching the street lights on and off in full-automated, semi-automated and manual method, is proficient to detect the faulty light in a system by measuring the threshold power level. This paper introduces a pragmatic implementation and the real time measurements with the projection of reduced power consumption. For the process of other paper works using IoT and controlling the intensity of lights, the internet connection has to be available continuously at night and so, the data processing method is more complicated as well as the maintenance cost is also high. Compared to these processes, this proposed model is evident to be more feasible.